\documentclass[12pt]{article}
\input epsf.tex
\newcommand{\beq}{\begin{equation}}
\newcommand{\eeq}{\end{equation}}
\newcommand{\beqa}{\begin{eqnarray}}
\newcommand{\eeqa}{\end{eqnarray}}
\newcommand{\beqar}{\begin{eqnarray*}}
\newcommand{\eeqar}{\end{eqnarray*}}

\newcommand{\inn}{\!\cdot\!}

\newcommand{\la}{\lambda}
\newcommand{\La}{\Lambda}

\newcommand{\z}{\zeta}

\newcommand{\eg}{{\it e.g.,}\ }
\newcommand{\ie}{{\it i.e.,}\ }
\newcommand{\labell}[1]{\label{#1}} %{\label{#1}} %
\newcommand{\reef}[1]{(\ref{#1})}
\newcommand\prt{\partial}

\newcommand\Tr{{\rm Tr}}
\newcommand\STr{{\rm STr}}
\begin{document}

\vspace*{1cm}

\begin{center}
{\bf \Large  On modified tachyon DBI action }
%Momentum expansion of the S-matrix element of \\
%four tachyons and one gauge field }
%On modified tachyon DBI action }

\vspace*{1cm}

{Mohammad R. Garousi}\\
\vspace*{0.2cm}
{ Department of Physics, Ferdowsi University of Mashhad,\\ P.O. Box 1436, Mashhad, Iran}
\\
\vspace*{0.1cm}
%and\\
{ School of Physics, 
 Institute for research in fundamental sciences (IPM), \\
 P.O. Box 19395-5531, Tehran, Iran. 
}\\
\vspace*{0.4cm}
\vspace{2cm}
ABSTRACT
\end{center}
Recently a modification of the tachyon DBI action has been proposed in which the tachyon carries the  internal CP matrix $\sigma_1$ and $\sigma_2$. In this paper, we find the momentum expansion of  the disk level S-matrix
element of four tachyons and one gauge field  in superstring
theory and  show that the first and second  order terms of the expansion  are
in perfect agreement with the  above tachyon DBI action. 

\vfill
\setcounter{page}{0}
\setcounter{footnote}{0}
\newpage

%\section{The idea} \label{intro}
\section{Introduction} \label{intro}
A natural mechanism for inflation in string theory is to have D-brane-anti-D-brane  separated along an extra dimension \cite{Dvali:1998pa,Dvali:2001fw,Burgess:2001fx,Kachru:2003sx}. In this model,  the brane separation plays the role of inflaton. When the separation is smaller than the string scale, the open string stretching between the branes become tachyonic at its ground state \cite{Banks:1995ch,Sen:1998ii,deAlwis:1999cg}. At this time the inflation stops and universe undergoes the reheating period when the energy of inflaton decays to the particles of the Standard Model \cite{Kofman:1997yn,Brandenberger:2007ca}. To understand this period,  one needs to know the effective action of brane-antibrane.  Even though the mass scale of the tachyon and the mass scale of the massive modes of the open string are the same order, there are various arguments that indicate there must be an effective action for brane-antibrane in which there are only tachyon and massless fields \cite{Sen:2004nf}. The effective action of brane-antibrane may be found from the  effective action of two non-BPS branes by $(-1)^{F_L}$ orbifolding \cite{Garousi:2004rd}.

The effective theory  of non-BPS branes has  two parts, \ie 
\beqa
S_{non-BPS}&=&S_{DBI}+S_{WZ}\labell{nonbps}\nonumber\eeqa
where $S_{DBI}/S_{WZ}$ must be an extension of the DBI/WZ action of BPS branes in which the tachyon modes of the non-BPS branes are included appropriately. There are various methods to study these actions. It has been shown in \cite{Sen:2002in,Sen:2002an} that the tachyon DBI action \cite{Sen:1999md,Garousi:2000tr,Bergshoeff:2000dq,Kluson:2000iy} can capture many properties of decay of non-BPS branes around the stable point of the tachyon potential. In \cite{Kennedy:1999nn}, using the S-matrix method, an extension for the  WZ action of non-BPS branes/brane-antibrane has been proposed  in which the gauge field  has been replaced by the superconnection. This action is consistent with the leading order terms of the momentum expansion of the S-matrix element of one RR and two tachyons \cite{Kennedy:1999nn}.  This form of action has been then confirmed by the BSFT method in \cite{Kraus:2000nj,Takayanagi:2000rz}. See \cite{Garousi:2007fk,Garousi:2008tn} for consistency of the WZ action with other S-matrix elements.  

However,  the BSFT and the S-matrix methods produce  different actions for the DBI part which may be related to each other by some field redefinition. The S-matrix method indicates that the leading order terms of the momentum expansion of any S-matrix element which involves two tachyon vertex operators are consistent with the usual tachyon DBI action \cite{Garousi:2000tr}. However, the momentum expansion of any S-matrix element that involves four or more tachyon vertex operators must be consistent with the modification of the tachyon DBI action in which the tachyon carries the Pauli matrices $\sigma_1$ and/or $\sigma_2$ \cite{Garousi:2008tn,Garousi:2008ze}. This action in flat background is 
\beqa
S_{DBI}&\sim&\int
d^{p+1}\sigma \STr\left(\frac{}{}V({ T_iT_i})\sqrt{1+\frac{1}{2}[T_i,T_j][T_j,T_i]}\right.\labell{nonab} \\
&&\qquad\qquad\left.
\times\sqrt{-\det(\eta_{ab}
+2\pi\alpha'F_{ab}+2\pi\alpha'D_a{ T_i}(Q^{-1})_{ij}D_b{ T_j})} \right)\,,\nonumber\eeqa  where $V({T_iT_i})=e^{\pi\alpha'm^2{ T_iT_i}/2}$ and  $m^2=-1/(2\alpha')$. In above action
\beqa
Q_{ij}&=&I\delta_{ij}-i[T_i,T_j]\eeqa
The subscripts $i,j=1,2$, \ie $T_1=T\sigma_1$, $T_2=T\sigma_2$. In above action  there is no sum over $i,j$. To fix these indices, one  must first  expand the square roots and then  choose two of indices to be $2$  and all others to be $1$. The trace in above equation
should be completely symmetric between all  matrices
of the form $F_{ab},D_a{ T_i}$, $[T_i,T_j]$ and individual
${ T_i}$ of the  potential $V(T_iT_i)$. This symmetric trace makes the above rule for fixing the indices $i,j$ to be unambitious. Around the stable point of the tachyon potential and for abelian gauge group, the above action reduces to the usual tachyon DBI action  with the potential $T^4V$ (see the Discussion section).

The above modification to the tachyon DBI action has been found in \cite{Garousi:2008tn} by studying the S-matrix element of one RR and three tachyons. The tachyon pole in this amplitude must be reproduced in field theory by considering appropriate four-tachyon couplings. The four-tachyon couplings in the usual tachyon DBI action can not  reproduce the tachyon pole of the S-matrix element. However, by assuming the tachyon in the action carries the Pauli matrices as above, one can produce the tachyon pole exactly \cite{Garousi:2008tn}. This rule is consistent with the observation that the open string states of non-BPS branes carry interal matrices \cite{Sen:1999mg}. To see this, we note that the tachyon of non-BPS branes in picture $(0)$ carries internal matrix $\sigma_1$ \cite{Sen:1999mg}. The picture changing operator, on the other hand, carries the internal matrix $\sigma_3$ \cite{DeSmet:2000je}, hence, the tachyon in picture $(-1)$ must carry internal matrix $\sigma_2$. Now, using the fact that the superghost charge of the disk level amplitude is $-2$, \ie two of the vertex operators must be in $(-1)$ picture and all other must be in $(0)$ picture, and using the observation that the momentum expansion of the string theory S-matrix element of tachyons in which two of the tachyon vertex operators to be in $(-1)$ picture and all other to be in $(0)$ picture is very similar to the momentum expansion of the massless transverse scalars \cite{Garousi:2008tn,Garousi:2008ze},  one finds the above rule for fixing the Pauli matrices in the modified tachyon DBI action \reef{nonab}.

 In this paper we would like to compare the action \reef{nonab} with the S-matrix element of four tachyons and one gauge field. This study can confirm the presence of covariant derivative in the action and more importantly this can confirm the presence of $Q_{ij}$ which does not appear in the four tachyon couplings studied in \cite{Garousi:2008tn}. So in the next section we write the S-matrix element by including the internal matrices in the vertex operators, and find the momentum expansion of the amplitude. The amplitude has massless/tachyon poles and contact terms at each order. In section 3, using the action \reef{nonab}, we reproduce exactly the massless/tachyon poles and the contact terms of the S-matrix element at the first and second orders.  Section 4 devoted to a brief discussion.

\section{Tachyon amplitude in superstring theory}

The S-matrix element of four tachyons and one gauge
field of N non-BPS D$_p$-branes in superstring theory may be  given by
the following correlation function: 
\beqa
\sum_{\rm non-cyclic}
\int dx_1dx_2dx_3dx_4dx_5~\left(\langle V_{-1}^{T}(x_1)V_{-1}^{T}(x_2)V_{0}^{T}(x_3)V_{0}^{T}(x_4)
V_{0}^{A}(x_5)\rangle \right) \labell{cor00}\eeqa where the vertex operators
for tachyons and gauge field are given by\footnote{Our conventions in string theory side set $\alpha'=2$.}
 \beqa V_0^{T}&=&(2ik\inn \psi)e^{2ik\inn X}\lambda\otimes\sigma_1 \nonumber\\
 V_{-1}^{T}&=&e^{2ik\inn X}e^{-\Phi} \lambda\otimes\sigma_2 \nonumber\\
  V_{0}^{A}&=&\xi_a(\prt X^a+2ik\inn\psi\psi^a)e^{2ik\inn X}\lambda\otimes I\nonumber \eeqa
where     $\xi_a$ is the polarization of gauge
field. The matrix
$\lambda$ is the external  Chan-Paton matrix and $\sigma_1,\,\sigma_2,\,I$ are the internal CP matrices.
In the above vertexes,  $k$ is the world volume momentum of the
open string states. The on-shell condition for tachyon is
$k^2=1/(2\alpha')$, and for gauge field is $\xi_a k^a=0$ and
$k^2=0$. 

Momentum expansion of the above S-matrix element should produce an effective field theory in which  the massless fields    carry identity internal matrix, because when tachyon is set to zero the effective action of non-BPS branes must be reduced to the effective action of BPS branes that has no internal CP matrix. Hence,  one finds that the momentum expansion of above amplitude  should have massless pole in $s_{12},\,s_{34}$, \ie $s_{12},s_{34}\rightarrow 0$.   In particular, it does not have massless pole in $s_{23}$, \ie  to have a massless pole in this channel, the gauge field should carry $\sigma_3$ which is forbidden. To have a S-matrix element whose momentum expansion  is consistent with the effective field theory, one has to add two other correlators in \reef{cor00}, \ie 
\beqa {\cal A}&\sim&\sum_{\rm non-cyclic}
\int dx_1dx_2dx_3dx_4dx_5~\left(\langle V_{-1}^{T}(x_1)V_{-1}^{T}(x_2)V_{0}^{T}(x_3)V_{0}^{T}(x_4)
V_{0}^{A}(x_5)\rangle  \labell{amp0}\right.\\
&&\left.+~\langle V_{-1}^{T}(x_1)V_{0}^{T}(x_2)V_{-1}^{T}(x_3)V_{0}^{T}(x_4)
V_{0}^{A}(x_5)\rangle \right.\nonumber\\
&&\left.+~\langle V_{-1}^{T}(x_1)V_{0}^{T}(x_2)V_{0}^{T}(x_3)V_{-1}^{T}(x_4)
V_{0}^{A}(x_5)\rangle \right)\nonumber\eeqa
The momentum expansion of the second correlator should have massless pole in $s_{13},\, s_{24}$, \ie $s_{13},s_{24}\rightarrow 0$, and the last correlator should have massless pole in $s_{14},\, s_{23}$, \ie $s_{14},s_{23}\rightarrow 0$. All above correlators should have  tachyonic pole in $s_{15}$, $s_{25}$, $s_{35}$ and $s_{45}$, and massive poles in all other channels. Note that the above correlators for different ways of distributing the superconformal ghost charge are equivalent when the internal CP factors are included \cite{Garousi:2008ze}, however, their momentum expansion are different.

Alternatively, the S-matrix element can be given by the following correlator: 
\beqa
 {\cal A}&\sim&\sum_{\rm non-cyclic}
\int dx_1dx_2dx_3dx_4dx_5~\left(\langle V_{0}^{T}(x_1)V_{0}^{T}(x_2)V_{0}^{T}(x_3)V_{0}^{T}(x_4)
V_{-2}^{A}(x_5)\rangle \right) \labell{amp01}\eeqa
In this case, the gauge field carries the identity internal matrix and all tachyon vertex operators carry $\sigma_1$. The momentum expansion of this amplitude has the same massless/tachyon poles as the amplitude \reef{amp0}. The amplitude is symmetric under interchanging the tachyons, so the expansion here should be around 
\beqa
(s_{12},s_{34})\rightarrow 0\quad +\quad (s_{13},s_{24})\rightarrow 0\quad +\quad (s_{14},s_{23})\rightarrow 0\labell{55}\eeqa
which is symmetric  under interchanging the tachyons. The first, second and the third term above  are corresponding to the momentum expansion of the first, second and the third correlator in \reef{amp0}, respectively.

The definition of Mandelstam variables is
\beqa
s_{ij}&=&-\alpha'(k_i+k_j)^2\eeqa
The number of independent kinematic factors in the scattering
amplitude of $n$ states is $\frac{n}{2}(n-3)$ \cite{ZK}. In the
present case, there are 5 independent kinematic factors. One may choose them  to be $s_{12},s_{23},s_{34},s_{45},s_{15}$. Using  conservation of
momentums and the on-shell conditions
$k_1^2=k_2^2=k_3^2=k_4^2=1/(2\alpha'),\,k_5^2=0$, one finds that
the other kinematic factors $s_{13},s_{14},s_{24},s_{25},s_{35}$
can be written in terms of the independent ones as \beqa
s_{13}&=&s_{45}-s_{12}-s_{23}-\frac{3}{2}\nonumber\\s_{14}&=&s_{23}-s_{15}-s_{45}-1
\nonumber\\s_{24}&=&s_{15}-s_{23}-s_{34}-\frac{3}{2}
\labell{cons1}\\s_{25}&=&s_{34}-s_{12}-s_{15}-1\nonumber\\s_{35}&=&s_{12}-s_{45}-s_{34}-1\nonumber\eeqa
One can  show that the Mandelstam variables satisfy the
following relation: \beqa
\sum_{i<j}s_{ij}&=&-6\labell{onshell}\eeqa

One should write  the momentum expansion of the above amplitude in terms of independent variables.  In \cite{Garousi:2007si}, it has been argued  that the momentum expansion of a S-matrix element should be, in general,  around $(k_i+k_j)^2\rightarrow 0$ and/or $k_i\inn k_j\rightarrow 0$ where $k_i$ is the momentum of $i$-th particle. The expansion $(k_i+k_j)^2\rightarrow 0$ should be corresponding to the massless poles. So, in the first corrlator in \reef{amp0} the expansion should be around 
\beqa
((k_1+k_2)^2,\, (k_3+k_4)^2)\rightarrow 0, & {\rm and}& (k_2\inn k_3,\,k_1\inn k_5,\, k_4\inn k_5)\rightarrow 0\labell{exp1}
\eeqa
In the second correlator in \reef{amp0}, since there is no massless pole for the independent variables the expansion should be around
\beqa
 (k_1\inn k_2,\,k_3\inn k_4,\,k_2\inn k_3,\,k_1\inn k_5,\, k_4\inn k_5)\rightarrow 0\labell{exp2}
\eeqa
 In the last correlator in \reef{amp0}, there is massless pole only in the $s_{23}$-channel so the expansion should be around 
\beqa
(k_2+k_3)^2,\, \rightarrow 0, & {\rm and}& (k_1\inn k_2,\,k_3\inn k_4,\,k_1\inn k_5,\, k_4\inn k_5)\rightarrow 0\labell{exp3}
\eeqa

We now try to calculate the correlators in \reef{amp0}. 
It has been shown in \cite{Garousi:2008ze}, that  when the internal CP factors are included, the S-matrix elements in different  arrangement of the picture of the vertex operators  are identical. So each of the terms in \reef{amp0}  should be the same as 
\beqa {\cal A'}&\sim&
\sum_{\rm non-cyclic}\int dx_1dx_2dx_3dx_4dx_5~\langle V_{0}^{T}(x_1)V_{0}^{T}(x_2)V_{0}^{T}(x_3)V_{-1}^{T}(x_4)
V_{-1}^{A}(x_5)\rangle \labell{amp00}\eeqa
The only difference is in their internal CP factors.  The CP factor for 12345 ordering in the first, second and the third term of \reef{amp0} are $\Tr(\sigma_2\sigma_2\sigma_1\sigma_1 I)=2$,  $\Tr(\sigma_2\sigma_1\sigma_2\sigma_1 I)=-2$ and $\Tr(\sigma_2\sigma_1\sigma_1\sigma_2 I)=2$, respectively. On the other hand the CP factor of the  amplitude ${\cal A'}$ is  $\Tr(\sigma_1\sigma_1\sigma_1\sigma_2\sigma_3)=-2i$. The  correlation function in ${\cal A'}$ for 12345 ordering has been calculated in \cite{BitaghsirFadafan:2006cj}. The result has three terms which are proportional to $k_1\inn\xi$, $k_2\inn\xi$ and $k_3\inn\xi$. The one which is proportional to $k_1\inn\xi$ is 
 \beqa {\cal A'} &=&-2\alpha'T_p
\Tr(\lambda^1\lambda^2\lambda^3\lambda^4\lambda^5)\Tr(\sigma_1\sigma_1\sigma_1\sigma_2\sigma_3)k_1\inn\xi \nonumber\\&&\times
(-s_{23}-1)\beta(-s_{12},-s_{23}-1)\beta(-s_{45}-\frac{1}{2},-s_{34})\nonumber\\&&\times
{}_{3}F_2(-s_{12},-s_{25}-\frac{1}{2},s_{15}-s_{23}-s_{34}-\frac{1}{2};-s_{12}-s_{23}-1,-s_{34}-s_{45}-\frac{1}{2};1)\nonumber\eeqa 
So the $ \Tr(\lambda^1\lambda^2\lambda^3\lambda^4\lambda^5) k_1\inn\xi $ part of the amplitude \reef{amp0} should be equal to 
\beqa {\cal A} &=&i2\alpha'T_p
\Tr(\lambda^1\lambda^2\lambda^3\lambda^4\lambda^5)(\Tr(\sigma_2\sigma_2\sigma_1\sigma_1 I)-  \Tr(\sigma_2\sigma_1\sigma_2\sigma_1 I)+\Tr(\sigma_2\sigma_1\sigma_1\sigma_2 I))k_1\inn\xi \nonumber\\&&\times
(-s_{23}-1)\beta(-s_{12},-s_{23}-1)\beta(-s_{45}-\frac{1}{2},-s_{34})\labell{amp}\\&&\times
{}_{3}F_2(-s_{12},-s_{25}-\frac{1}{2},s_{15}-s_{23}-s_{34}-\frac{1}{2};-s_{12}-s_{23}-1,-s_{34}-s_{45}-\frac{1}{2};1)\nonumber\eeqa 
The first, second and the third term above correspond to the first, second and the third correlator in \reef{amp0}, respectively. Note that   the above amplitude is  similar to the amplitude of four massless transverse scalars and one gauge field, in particular, the appearance of minus sign in the second term above \cite{BitaghsirFadafan:2006cj}. There is similar symmetry in the S-matrix element of four tachyons \cite{Garousi:2008tn,Garousi:2008ze}. 

From  the poles of the Beta and
Hypergeometric functions, one realizes that
the amplitude has  tachyon, massless and infinite tower of massive
poles. The  tachyon and massless  poles should be
reproduced by effective  field theory. The
 momentum expansion for Beta and the
Hypergeometric functions  must keep only the tachyon and the
massless poles 
and expands  all other poles to produce infinite number of contact terms which are ordered in terms of the momenta of the external states.

The momentum expansion in equations \reef{exp1}, \reef{exp2} and \reef{exp3} in terms of the independent Mandelstam variables are $(s_{12},s_{34}\rightarrow 0,\, s_{23}\rightarrow -1,\,s_{45},s_{15}\rightarrow -1/2)$, $(s_{23},s_{12},s_{34}\rightarrow -1,\,s_{45},s_{15}\rightarrow -1/2)$ and $(s_{23}\rightarrow 0,\, s_{12},s_{34}\rightarrow -1,\,s_{45},s_{15}\rightarrow -1/2)$, respectively. One can easily check that these limits are consistent with the constraint \reef{onshell}. 
%These variables are not appropriate choice for the independent Mandelstam variables because, for instance, $s_{23}$ once goes to $-1$ and once goes to $0$. We choose another set of independent variables as $s_{15},s_{25},s_{35},s_{45},s_{24}$. 
Using the package HypExp \cite{Huber:2005yg} for expanding the Hypergeometric functions, one can expand  the amplitude \reef{amp} around the above points. The result is:
\beqa {\cal
A}^{TTTTA} &\!\!\!\!=\!\!\!\!&4i\alpha'T_p
Tr(\lambda^1\lambda^2\lambda^3\lambda^4\lambda^5)\xi_5 \cdot
k_1\labell{Atttta}\\&& \left[
  \left( \frac{s_{23}+1}{(s_{45}+\frac{1}{2}) s_{12}}
    + \frac{s_{23}+1}{(s_{15}+\frac{1}{2}) s_{34}}
 + \frac{s_{23}+1}{s_{12} s_{34}}+ \frac{s_{12}+1}{s_{23} (s_{45}+\frac{1}{2})}
  + \frac{s_{34}+1}{s_{23}(s_{15}+\frac{1}{2})}\right.\right.\nonumber\\&&
 +\left. \frac{3}{(s_{15}+\frac{1}{2})}+\frac{3}{(s_{45}+\frac{1}{2})}-\frac{1}{s_{23}}
\right) \nonumber\\&&
-\zeta(2)\left(\frac{s_{12}^2}{(s_{45}+\frac{1}{2})}+\frac{s_{23}^2}{(s_{45}+\frac{1}{2})}+
\frac{3s_{12}s_{23}}{(s_{45}+\frac{1}{2})}+\frac{4s_{12}}{(s_{45}+\frac{1}{2})}
+\frac{4s_{23}}{(s_{45}+\frac{1}{2})}+\frac{3}{(s_{45}+\frac{1}{2})}\right.\nonumber\\&&
+\frac{s_{23}^2}{(s_{15}+\frac{1}{2})}+\frac{s_{34}^2}{(s_{15}+\frac{1}{2})}+
\frac{3s_{23}s_{34}}{(s_{15}+\frac{1}{2})}+
\frac{4s_{23}}{(s_{15}+\frac{1}{2})}+\frac{4s_{34}}{(s_{15}+\frac{1}{2})}+\frac{3}{(s_{1
5}+\frac{1}{2})}\nonumber\\&&
+\frac{s_{23}s_{12}}{s_{34}}+\frac{s_{23}(s_{15}+\frac{1}{2})}{s_{34}}+\frac{s_{12}}{s_{34}}
+\frac{(s_{15}+\frac{1}{2})}{s_{34}}
+\frac{s_{23}s_{34}}{s_{12}}+\frac{s_{23}(s_{45}+\frac{1}{2})}{s_{12}}+\frac{s_{34}}{s_{12}}
\nonumber\\&&+\frac{(s_{45}+\frac{1}{2})}{s_{12}}+\frac
{s_{12}(s_{15}+\frac{1}{2})}{s_{23}}+\frac
 {s_{34}(s_{45}+\frac{1}{2})}{s_{23}} +\frac
{(s_{45}+\frac{1}{2})}{s_{23}}-\frac{(s_{15}+\frac{1}{2})(s_{45}+\frac{1}{2})}{s_{23}}\nonumber\\&&
\left.\left.+\frac{(s_{15}+\frac{1}{2})}{s_{23}}
 -s_{12}-s_{34}+3s_{15}+3s_{45}-5s_{23}-3\right)+\cdots \right]\nonumber\eeqa
As we have anticipated, the above  expansion keeps  the tachyon
and the massless poles of the amplitude \reef{amp} and expands all other
poles. Obviously, one can rewrite the expansion in terms of the momenta of the external states, because the expansion \reef{exp1}, \reef{exp2} and \reef{exp3} are in terms of the momenta of the external states. The terms in the second and third lines above are $1/\alpha'$, the terms proportional to $\z(2)$ are $\alpha'$ order, and dots refers to the higher order of $\alpha'$. In terms of $s_{ij}$, the above  terms  have different $\alpha'$ order.  This indicates that the effective field theory that should reproduce them must have  couplings with  different $\alpha'$ order.   In the next
section, we shall show that the terms in the second and the third  lines
above are reproduced by  the terms in
the first line of \reef{expandL}, and the terms proportional to
$\z(2)$ are reproduced by all other terms in \reef{expandL} which have
obviously different $\alpha'$ order.

\section{Amplitude in effective field theory}
%%%%%%%%%%%%%%%%%%%%%%%%%%%%
%%%%%%%%%%%%%%%%%%%%%%
Now we would like to compare the above terms of the expansion of the S-matrix element with the modified tachyon DBI action \reef{nonab}.
One can expand the square roots in the action \reef{nonab} to find various couplings. The terms of the  expansion which has
contribution to the S-matrix element of one gauge field and four
tachyons are the following: \beqa {\cal
L}&=&-\frac{T_p}{2}\Tr\left((\pi\alpha')m^2T_2T_2+(\pi\alpha')D_aT_2D^aT_2-(\pi\alpha')^2F_{ab}F^{ba}+[T_1,T_2][T_2,T_1]/4\right)\nonumber\\
&&-\frac{T_p}{2}(2\pi\alpha')^2S\Tr\left(\frac{m^4}{8}T_1T_1T_2T_2+\frac{m^2}{4}T_1T_1D_aT_2D^aT_2\right.\labell{expandL}\\
&&\qquad\qquad\qquad\qquad\left.-\frac{1}{4}(D_aT_1D_bT_1D^bT_2D^aT_2)+\frac{1}{8}(D_aT_1D^aT_1)(D_bT_2D^bT_2)\right)\nonumber\\
&&-\frac{T_p}{2}(2\pi\alpha')^2S\Tr\left(\frac{i}{2}D_aT_1[T_1,T_2]D_bT_2F^{ba}\right)\nonumber\\
&&-\frac{T_p}{2}\frac{(2\pi\alpha')^3}{2}S\Tr\left(D^aT_2D_bT_2F^{bc}F_{ca}
-\frac{1}{4}D^aT_2D_aT_2F^{bc}F_{cb}-\frac{m^2}{4}T_2T_2F^{ab}F_{ba}\right)\nonumber\eeqa
 Note that the above terms are not ordered in terms of power of
$\alpha'$. This  is consistent with the momentum expansion of the S-matrix element \reef{Atttta}.  We shall show that the terms in the first line above
which we call them kinetic order terms, reproduce the first
leading order terms in \reef{Atttta}, and the other terms  reproduce
the  terms in \reef{Atttta} which are proportional to  $\z(2)$.

\subsection{Kinetic  order terms}
Performing the trace over the internal CP matrices in the first line of \reef{expandL}, one finds
\beqa
-T_p\Tr\left((\pi\alpha')m^2TT+(\pi\alpha')D_aTD^aT-(\pi\alpha')^2F_{ab}F^{ba}+T^4\right)\labell{kinL}
\eeqa
The non-abelian field strength and covariant
derivative of tachyon are, respectively,
 \beqa F^{ab}=
\partial^aA^b-\partial^bA^a-i[A^a,A^b],~~~~
D_aT_j=\partial_aT_j-i[A_a,T_j] \nonumber\eeqa   Using the  couplings \reef{kinL}, one can calculate the Feynman amplitude corresponding to the massless and tachyon poles of the string theory S-matrix element \reef{Atttta}. Similar calculation has been done in \cite{BitaghsirFadafan:2006cj}. However, the last term in the above equation  does not appear in the couplings considered in \cite{BitaghsirFadafan:2006cj}.  The Feynman amplitude at the kinetic order term which results from this term is given by
 \beqa
&&V^{\alpha}(T_4A_5T)G_{\alpha\beta}(T)V^{\beta}(TT_1T_2T_3)+V^{\alpha}(T_1A_5T)G_{\alpha\beta}(T)V^{\beta}(TT_2T_3T_4)\nonumber\\
 &&=4i\alpha'T_p k_1\inn\xi_5\Tr(\la_1\la_2\la_3\la_4\la_5)\left(\frac{2}{s_{45}+\frac{1}{2}}+\frac{2}{s_{15}+\frac{1}{2}}\right)+\cdots
 \eeqa
where we have kept only the terms that are proportional to $k_1\inn\xi_5\Tr(\la_1\la_2\la_3\la_4\la_5)$. Adding them to the result in \cite{BitaghsirFadafan:2006cj} (eq.(23) of \cite{BitaghsirFadafan:2006cj}), one finds  
 \beqa {\cal
A}_k^{TTTTA}&=&(i\alpha'T_p)k_1\cdot\xi_5\left(\frac{12}{s_{45}+\frac{1}{2}}+\frac{12}{s_{15}+\frac{1}{2}}-\frac{4}{s_{23}}
+\frac{4s_{23}}{(s_{45}+\frac{1}{2})s_{12}}+\frac{4s_{23}}{(s_{15}+\frac{1}{2})s_{34}}\right.\nonumber\\&&+
\frac{4s_{34}}{(s_{15}+\frac{1}{2})s_{23}}+\frac{4s_{12}}{(s_{45}+\frac{1}{2})s_{23}}+
\frac{4s_{23}}{s_{34}s_{12}}+\frac{4}{(s_{45}+\frac{1}{2})s_{12}}+\frac{4}{(s_{15}+\frac{1}{2})s_{34}}
\nonumber\\&&\left.+\frac{4}{(s_{15}+\frac{1}{2})s_{23}}+\frac{4}{(s_{45}+\frac{1}{2})s_{23}}+
\frac{4}{s_{34}s_{12}}\right)\Tr(\la^{1}\la^{2}\la^{3}\la^{4}\la^{5})+\cdots\labell{Ak}\eeqa
where dots refers to the terms which have  coefficients other than $\Tr(\la^{1}\la^{2}\la^{3}\la^{4}\la^{5})k_1\cdot\xi_5$. 
This is exactly the leading order terms of the momentum
expansion of the S-matrix element in the string theory side
\reef{Atttta}.   The next
 order terms in \reef{Atttta} are proportional to $\z(2)$.
We now turn to the Feynman amplitudes in field theory that are
proportional to $\z(2)$.

\subsection{$\zeta(2)$ order  terms}

There are three   Feynman amplitudes in the field theory at this order. One amplitude is produced by one
vertex from the kinetic term of tachyon, one tachyon propagator and one vertex from four-tachyon couplings in \reef{expandL}. The second amplitude is produced by one vertex from the kinetic term of tachyon, one gauge field propagator and one vertex from the two-gauge-two-tachyon couplings in \reef{expandL}. And the last amplitude is the contact term of one-gauge-four-tachyon couplings in \reef{expandL}. We write them, respectively,  as  
\beqa {\cal
A}^{TTTTA}_{\z(2)}&=&A_{\z(2)}+A'_{\z(2)}+A''_{\z(2)}\nonumber\eeqa  The Feynman amplitude $A_{\z(2)}(T_1T_2T_3T_4A_5)$ 
is given by \beqa A_{\z(2)}(T_1T_2T_3T_4A_5)&=&{
V}_{\alpha}(T_1T_2T_3T)G_{\alpha\beta}(T){
V}_{\beta}(TT_4A_5)\labell{amp4}\eeqa The vertex of four-tachyons
should be read from the different terms in the second line of
\reef{expandL}. To do this, one should first perform the symmetric
trace and perform the trace over the internal CP matrices, \ie
\beqa {\cal
L}^{TTTT}&=&-\frac{T_p}{2}(2\pi\alpha')^2\Tr\left(\frac{m^4}{24}T^4+\frac{m^2}{4}\left(\frac{2}{3}TTD_aTD^aT
-\frac{1}{3}TD_aTTD^aT\right)\right.\nonumber\\&&
+\left.\frac{1}{24}\left(2D_aTD^aTD_bTD^bT-3D_aTD_bTD^aTD^bT\frac{}{}\right)\right)\labell{Ltttt}
\eeqa 
Now we can read the vertex of three on-shell and
one off-shell tachyons from the above couplings. The
vertex of four-tachyon  contains terms that have group factors
$\Tr(\la^1\la^2\la^3\La^{\alpha})$, $\Tr(\la^2\la^1\la^3\La^{\alpha})$,
$\Tr(\la^2\la^3\la^1\La^{\alpha})$, $\Tr(\la^3\la^2\la^1\La^{\alpha})$,
$\Tr(\la^1\la^3\la^2\La^{\alpha})$, and
$\Tr(\la^3\la^1\la^2\La^{\alpha})$. After replacing them in
\reef{amp4}, only the first factor will produce the desired
ordering $\Tr(\la^1\la^2\la^3\la^4\la^5)$. So, we  consider only the
terms in the vertex that have factor
$\Tr(\la^1\la^2\la^3\La^{\alpha})$. 
%On the other hand, since  the tachyon in  the kinetic term is $T_2$,  the off-shell tachyon in the four-tachyon vertex must be $T_2$. 
Hence, the vertex is \beqa {
V}_{\alpha}(T_1T_2T_3T)&=&-T_pi(2\pi\alpha')^2
\Tr(\lambda^{1}\lambda^{2}\lambda^{3}\Lambda^{\alpha})\nonumber\\&&
\times\left(\frac{m^4}{6}-\frac{m^2}{6} (k_3\inn p+k_1\inn p+k_1\inn k_2+k_2\inn k_3-k_2\inn p-k_1\inn k_3)\right.\nonumber\\
&&\left.+\frac{1}{6}(k_1\inn k_2 k_3\inn p+ k_2\inn k_3 k_1\inn p-3k_1\inn k_3 k_2\inn p)\right)+\cdots\labell{ver2}\eeqa where $p$
is momentum of off-shell tachyon.  Replacing
this and the other vertex and propagator in \reef{amp4}, one finds
\beqa
A_{\z(2)}(T_1T_2T_3T_4A_5)&=&2i\alpha'T_p\z(2)\Tr(\lambda^{1}\lambda^{2}\lambda^{3}\lambda^{4}\lambda^{5})
 k_1\cdot \xi_5\nonumber\\&&\times
 \left(-\frac{2s_{12}^2}{s_{45}+\frac{1}{2}}-\frac{6s_{12}s_{23}}{s_{45}+\frac{1}{2}}
 -\frac{2s_{23}^2}{s_{45}+\frac{1}{2}}-\frac{7s_{12}}{s_{45}+\frac{1}{2}}
 -\frac{7s_{23}}{s_{45}+\frac{1}{2}}-\frac{4}{s_{45}+\frac{1}{2}}\right.\nonumber\\&&\left.
+\frac{2s_{45}s_{12}}{s_{45}+\frac{1}{2}}+\frac{2s_{45}s_{23}}{s_{45}+\frac{1}{2}}+\frac{4s_{45}}{s_{45}+\frac{1}{2}}\right)+\cdots\labell{amp6}\eeqa
where  we have  used relation \reef{cons1} and $\pi^2/6=\z(2)$.

 The other distinct  diagram that produces the
ordering
 $\Tr(\lambda^{1}\lambda^{2}\lambda^{3}\lambda^{4}\lambda^{5})$
 is
   \beqa A_{\z(2)}(T_2T_3T_4T_1A_5)&=&2i \alpha'
 T_p\z(2)\Tr(\lambda^{1}\lambda^{2}\lambda^{3}\lambda^{4}\lambda^{5})
 k_1\cdot \xi_5\nonumber\\&&
 \left(-\frac{2s_{23}^2}{s_{15}+\frac{1}{2}}-\frac{6s_{34}s_{23}}{s_{15}+\frac{1}{2}}
 -\frac{2s_{34}^2}{s_{15}+\frac{1}{2}}-\frac{7s_{23}}{s_{15}+\frac{1}{2}}
 -\frac{7s_{34}}{s_{15}+\frac{1}{2}}-\frac{4}{s_{15}+\frac{1}{2}}\right.\nonumber\\&&\left.
 +\frac{2s_{15}s_{23}}{s_{15}+\frac{1}{2}}+\frac{2s_{15}s_{34}}{s_{15}+\frac{1}{2}}
 +\frac{4s_{15}}{s_{15}+\frac{1}{2}}\right)+\cdots
\labell{amp7}\eeqa where again dots refer to the terms that have
coefficient $k_2\cdot\xi_5$, $k_3\cdot\xi_5$, and have group
factor other than
$\Tr(\lambda^{1}\lambda^{2}\lambda^{3}\lambda^{4}\lambda^{5})$.

We now consider the amplitude $A'_{\z(2)}$.  The amplitude $A'_{\z(2)}(A_5T_1T_2T_3T_4)$ is given
by \beqa A'_{\z(2)}(A_5T_1T_2T_3T_4)&=&{
V}^a_{\alpha}(A_5T_1T_2A)G^{ab}_{\alpha\beta}(A)
V^b_{\beta}(AT_3T_4)\labell{amp5}\eeqa where the vertex of two-gauge-two-tachyon  should be read from the terms in the
last line of \reef{expandL}. There are three different amplitudes here that have been calculated in \cite{BitaghsirFadafan:2006cj}, \ie
 
 \beqa A'_{\z(2)}(A_5T_1T_2T_3T_4)&=&i\alpha'T_p\z(2) k_1\cdot
\xi_5\Tr(\lambda^{1}\lambda^{2}\lambda^{3}\lambda^{4}\lambda^{5})
\nonumber\\&&\times\left(-\frac{4s_{23}(s_{15}+\frac{1}{2})}{s_{34}}-\frac{4s_{12}s_{23}}{s_{34}}
-\frac{4(s_{15}+\frac{1}{2})}{s_{34}}-\frac{4s_{12}}{s_{34}}\right.
\nonumber\\&&\left.
-2(s_{15}+\frac{1}{2})-2s_{12}+2s_{34}+4s_{23}+4\right)+\cdots\labell{amp8}\\ A'_{\z(2)}(T_3T_4A_5T_1T_2)&=&{
V}^a_{\alpha}(T_3T_4A_5A)G^{ab}_{\alpha\beta}(A)
V^b_{\beta}(AT_1T_2)\nonumber\\
&=&i\alpha'T_9\z(2) k_1\cdot
\xi_5\Tr(\lambda^{1}\lambda^{2}\lambda^{3}\lambda^{4}\lambda^{5})
\nonumber\\&&\times\left(-\frac{4s_{23}(s_{45}+\frac{1}{2})}{s_{12}}-\frac{4s_{23}s_{34}}{s_{12}}
-\frac{4(s_{45}+\frac{1}{2})}{s_{12}}
-\frac{4s_{34}}{s_{12}}\right.\nonumber\\&&\left.
-2(s_{45}+\frac{1}{2})+2s_{12}+4s_{23}-2s_{34}+4\right)+\cdots\labell{amp9}\\
 A'_{\z(2)}(T_4A_5T_1T_2T_3)&=&{
V}^a_{\alpha}(T_4A_5T_1A)G^{ab}_{\alpha\beta}(A){
V}^b_{\beta}(AT_2T_3)\nonumber\\
&=&i\alpha'T_9\z(2)k_1\cdot
\xi_5Tr(\lambda^{1}\lambda^{2}\lambda^{3}\lambda^{4}\lambda^{5})\nonumber
\\&&\times\left(\frac{4(s_{15}+\frac{1}{2})(s_{45}+\frac{1}{2})}{s_{23}}-\frac{4(s_{15}+\frac{1}{2})s_{12}}{s_{23}}
-\frac{4(s_{45}+\frac{1}{2})s_{34}}{s_{23}}\right.\labell{amp10}\\&&\left.-\frac{4(s_{15}+\frac{1}{2})}{s_{23}}
-\frac{4(s_{45}+\frac{1}{2})}{s_{23}}-2(s_{15}+\frac{1}{2})-2(s_{45}+\frac{1}{2})\right)+\cdots\nonumber\eeqa

Finally, we consider the amplitude ${ A}''_{\z(2)}$. The couplings in \reef{Ltttt} produce the following contact terms of four tachyons and one gauge field:
\beqa
 A''_{\z(2)}(DTDTDTDT)&=&4i\alpha'T_p\z(2) k_1\cdot
\xi_5\Tr(\lambda^{1}\lambda^{2}\lambda^{3}\lambda^{4}\lambda^{5})s_{23}
\eeqa
Other contact terms are coming from coupling in the fourth line of \reef{expandL}. In this term, the commutator $[T_1,T_2]$ is coming from $(Q^{-1})_{ij}$ in \reef{nonab}. Writing the symmetric trace in terms of ordinary trace and performing the trace over the internal CP matrices, one finds
\beqa
\frac{i}{3}T_p(2\pi\alpha')^2\Tr(D_aTD_bT TTF^{ba}+D_aTD_bTF^{ba}TT-D_aT TTD_bTF^{ba})\nonumber\eeqa
The above couplings  produce the following contact terms:
\beqa
{ A}''_{\z(2)}(DTDTTTF)&=&-8i\alpha'T_p\z(2) k_1\cdot
\xi_5\Tr(\lambda^{1}\lambda^{2}\lambda^{3}\lambda^{4}\lambda^{5})\left(1+s_{15}+s_{45}
\right)
\eeqa

Now comparing  the field theory amplitudes with the string theory amplitude \reef{Atttta}, one finds exact agreement, \ie  
\beqa {\cal A}^{TTTTA}-{\cal A}_k^{TTTTA}-{\cal
A}_{\z(2)}^{TTTTA}&=&0+\cdots \labell{left}\eeqa where dots refer to the terms with
coefficients $\z(3)$ which are $(\alpha')^2$ order, $\z(4)$ which are $(\alpha')^3$ order, and so on. The above consistency, in particular, confirms the presence of $(Q^{-1})_{ij}$ in the tachyon action \reef{nonab}.   
This ends our illustration of consistency between  the momentum 
expansion of  the S-matrix element \reef{Atttta}  and the non-abelian
tachyon DBI action \reef{nonab}.

\section{Discussion}

In this paper we have shown that the leading order terms , and  next to the leading order terms of the momentum expansion of the S-matrix element of four tachyons and one gauge field are reproduced exactly by the modified tachyon DBI action that recently has been proposed in \cite{Garousi:2008tn}. The next order terms in the expansion \reef{Atttta}  are $(\alpha')^2$ order.  They contains massless/tachyon poles
and contact terms. The higher derivative corrections  in general have field redefinition freedom \cite{AAT,AAT1}, so one may choose this freedom to relate them   to  the couplings in field theory which have second  derivative of $T$, \ie
couplings that include $\prt\prt T$. Similarly, the $(\alpha')^3$ terms of \reef{Atttta} may be related to the couplings that include $\prt\prt\prt T$, and so on. It would be interesting to finds these higher derivative terms explicitly.
%In this case, the modified tachyon DBI action is effective action of N non-BPS D-branes when the second derivative of tachyon is zero.

The couplings in \reef{expandL} have on-shell ambiguity/freedom, \ie $m^2T\sim D_aD^aT$. This ambiguity/freedom  can not be fixed even by  studying  the S-matrix element   in which the tachyon appears as off-shell in the tachyon pole. If one replaces $T$ with $D_aD^aT$ it does not change the tachyon poles but produces an extra contact terms. These contact terms however  cancels the contact terms that are resulted from replacing $T$ with $DDT$.  For example,  one may rewrite  the first term of \reef{Ltttt} as
\beqa
&&-T_p(2\pi\alpha')^2\Tr\left(\frac{m^4}{24}T^4\right)\rightarrow -T_p(2\pi\alpha')^2\Tr\left(\alpha\frac{m^4}{24}T^4+\beta\frac{m^2}{24}TTTD_aD^aT\right.\nonumber\\
&&\left.+\frac{\lambda}{24}TTTD_aD_bD^bD^aT +\frac{\sigma}{24}TTTD_aD_bD^aD^bT +\frac{\gamma}{24}TTTD_aD^aD_bD^bT\right)\labell{rewrite}
\eeqa
where  $\alpha+\beta+\gamma+\lambda+\sigma=1$. The  couplings on the right hand side reproduce exactly the same S-matrix element as the coupling on the left hand side. To see this, we note that the right hand side  gives the following  one-gauge-four-tachyon contact terms:
\beqa
i T_p(2\pi\alpha')^2 k_1\cdot
\xi_5\Tr(\lambda^{1}\lambda^{2}\lambda^{3}\lambda^{4}\lambda^{5})\left(\frac{(\gamma+\lambda+\sigma)}{24}(1-s_{15}-s_{45})-\beta\frac{m^2}{6}\right)\labell{con3}
\eeqa
However, these couplings  change also the vertex \reef{ver2} and so modifies  the amplitudes \reef{amp6} and \reef{amp7}. They gives the following extra contact terms: 
\beqa
2i\alpha'T_p\z(2)\Tr(\lambda^{1}\lambda^{2}\lambda^{3}\lambda^{4}\lambda^{5})
 k_1\cdot \xi_5\left(-\beta+(\gamma+\lambda+\sigma)(s_{45}+s_{15}-1)\right)
 \eeqa
They cancel exactly the contact terms in \reef{con3}. Hence, one has freedom to choose the constants $\alpha,\,\beta,\,\gamma,\,\lambda,\,\sigma$. The S-matrix method can only fix  $\alpha+\beta+\gamma+\lambda+\sigma=1$. Assuming the effective field theory at the DBI  order has no couplings which has $D_aD^aT$ terms, fixes the constants to  $\beta=\gamma=\lambda=\sigma=0$ which is consistent with the tachyon DBI action. 
%On the other hand, the choice $\alpha=0$ removes the coupling to the higher derivative terms which can be ignored for slowly varying tachyon field.

The modified tachyon DBI action \reef{nonab} has couplings which are different from the usual tachyon DBI action. This action  has been found from studying the S-matrix element around the unstable point of the tachyon potential, \ie $T=0$. One may extrapolate the action to the stable point of the tachyon potential, \ie $T\rightarrow\infty$. Around this point the action reduces to the usual tachyon DBI action. To see this, we note the following:
\beqa
\sqrt{1+\frac{1}{2}[T_i,T_j][T_j,T_i]}&=&1+\frac{1}{4}[T_1,T_2][T_2,T_1]-\frac{1}{32}[T_1,T_2][T_2,T_1][T_1,T_2][T_2,T_1]+\cdots\nonumber\\
&=&1+\frac{1}{4}[T_1,T_2][T_2,T_1]
\eeqa
where in the second line we have used  the prescription given for the modified tachyon DBI action \reef{nonab} that only two of the tachyons must be $T_2$. All  terms of the square root in the second line of \reef{nonab} and the terms of tachyon potential $V(T)$  that multiply the second term above must have $T_1$. So the square root in the second line of \reef{nonab} that multiplied the second term above is the usual tachyon DBI action. On the other hand, around $T\rightarrow\infty$ only the second term in the above expansion is important. Therefore, the action \reef{nonab} reduces to the usual tachyon DBI action with potential $T^4V(T^2)$. This potential goes to zero at $T\rightarrow\infty$ as expected from the tachyon condensation \cite{Sen:1999md}.

We have found the momentum expansion of the amplitude \reef{amp0} by finding the massless poles of each correlator, \ie in the first correlator $(k_1+k_2)^2,(k_3+k_4)^2\rightarrow 0$, in the second correlator $(k_1+k_3)^2,(k_2+k_4)^2\rightarrow 0$ and in the last correlator $(k_1+k_4)^2,(k_2+k_3)^2\rightarrow 0$. All other Mandelstam variables go as $k_i\inn k_j\rightarrow 0$. The momentum expansion of the amplitude \reef{amp01} is similar. The correlator here  can have massless pole in all $s_{12},s_{34},s_{13},s_{24},s_{14},s_{23}$ channels. However, the on-shell constraint \reef{cons1} does not allow the correlator to have massless pole in these channels at the same time. In this case, one has to send once $(k_1+k_2)^2,(k_3+k_4)^2\rightarrow 0$, once  $(k_1+k_3)^2,(k_2+k_4)^2\rightarrow 0$ and once $(k_1+k_4)^2,(k_2+k_3)^2\rightarrow 0$ as in \reef{55}. This can easily be extended to the n-point function. For example, the S-matrix element of six tachyons is given by 
\beqa
 \sum_{\rm non-cyclic}
\int dx_1dx_2dx_3dx_4dx_5dx_6~\left(\langle V_{0}^{T}(x_1)V_{0}^{T}(x_2)V_{0}^{T}(x_3)V_{0}^{T}(x_4)
V_{0}^{T}(x_5)V_{-2}^T(x_6)\rangle \right) \labell{6T}\eeqa
The result must be symmetric under interchanging the tachyons. The momentum expansion of this amplitude is  the following:
\beqa
&&(s_{12},s_{34},s_{56})\rightarrow 0 +(s_{12},s_{35},s_{46})\rightarrow 0+(s_{12},s_{36},s_{45})\rightarrow 0+\nonumber\\
&&(s_{13},s_{24},s_{56})\rightarrow 0 +(s_{13},s_{25},s_{46})\rightarrow 0+(s_{13},s_{26},s_{45})\rightarrow 0+\nonumber\\
&&(s_{14},s_{23},s_{56})\rightarrow 0 +(s_{14},s_{25},s_{36})\rightarrow 0+(s_{14},s_{26},s_{35})\rightarrow 0+\nonumber\\
&&(s_{15},s_{23},s_{46})\rightarrow 0 +(s_{15},s_{24},s_{36})\rightarrow 0+(s_{15},s_{26},s_{34})\rightarrow 0+\nonumber\\
&&(s_{16},s_{23},s_{45})\rightarrow 0 +(s_{16},s_{24},s_{35})\rightarrow 0+(s_{16},s_{25},s_{34})\rightarrow 0\labell{Mand00}
\eeqa
which  is symmetric under interchanging the tachyons. In each case all other Mandelstam variables go as $k_i\inn k_j\rightarrow 0$.  Of course, not all the Mandelstam variables are independent. In each case it it is easy to find the expansion in terms of the independent variables, \eg see the expansions in \reef{exp1}, \reef{exp2} and \reef{exp3}. 

The internal CP matrix for all the vertex operator in the above amplitude is $\sigma_1$. Alternatively, the amplitude can be written as
\beqa
 \sum_{\rm non-cyclic}&&
\int dx_1dx_2dx_3dx_4dx_5dx_6~\left(\langle V_{-1}^{T}(x_1)V_{-1}^{T}(x_2)V_{0}^{T}(x_3)V_{0}^{T}(x_4)
V_{0}^{T}(x_5)V_{0}^T(x_6)\rangle +\right.\nonumber\\
&&\left.\qquad\qquad\qquad\langle V_{-1}^{T}(x_1)V_{0}^{T}(x_2)V_{-1}^{T}(x_3)V_{0}^{T}(x_4)
V_{0}^{T}(x_5)V_{0}^T(x_6)\rangle +\right.\nonumber\\
&&\left.\qquad\qquad\qquad\langle V_{-1}^{T}(x_1)V_{0}^{T}(x_2)V_{0}^{T}(x_3)V_{-1}^{T}(x_4)
V_{0}^{T}(x_5)V_{0}^T(x_6)\rangle +\right.\nonumber\\
&&\left.\qquad\qquad\qquad\langle V_{-1}^{T}(x_1)V_{0}^{T}(x_2)V_{0}^{T}(x_3)V_{0}^{T}(x_4)
V_{-1}^{T}(x_5)V_{0}^T(x_6)\rangle +\right.\nonumber\\
&&\left.\qquad\qquad\qquad\langle V_{-1}^{T}(x_1)V_{0}^{T}(x_2)V_{0}^{T}(x_3)V_{0}^{T}(x_4)
V_{0}^{T}(x_5)V_{-1}^T(x_6)\rangle \right) \nonumber\eeqa
For the first correlator, the Mandelstam variables go as in the first line of \reef{Mand00}, for the second correlator, the Mandelstam variables go as in the second  line of \reef{Mand00}, and so on. If one performs the correltors in the above amplitude and then use the above expansion, one should find consistency between the leading order terms of the expansion and the six-tachyon couplings in \reef{nonab}. It would be interesting to perform this calculation.

{\bf Acknowledgements}: This work was supported by a grant from Ferdowsi University of Mashhad. 
\newpage

%%%%%%%%%%%%%%%%%%%%%%%%%%%%%%%%%%%%%%%%
%%%%%%%%%%%%%%%%%%%%%%%%%%%%%%%%%%%%%%%
%\beqa -\frac{\hbar^2}{2\mu}\frac{\partial^2}{\partial x^2} \eeqa
%\newpage

\end{document}